\documentclass{article}
\usepackage{ijcai25}

\usepackage{times}
\usepackage{soul}
\usepackage{url}
\usepackage[hidelinks]{hyperref}
\usepackage[utf8]{inputenc}
\usepackage[small]{caption}
\usepackage{graphicx}
\usepackage{amsmath}
\usepackage{amsthm}
\usepackage{amssymb}
\usepackage{booktabs}
\usepackage{algorithm}
\usepackage{algorithmic}
\usepackage[switch]{lineno}
\usepackage{subcaption}
\usepackage{multirow}
\usepackage{xcolor}
\usepackage{enumitem}
\usepackage{bm}
\usepackage{microtype}

\DeclareMathOperator*{\argmax}{arg\,max}

\newtheorem{defn}{Definition}

\title{Modeling Behavioral Preferences of Cyber Adversaries Using Inverse Reinforcement Learning}

\author{
Aditya Shinde$^1$
\and
Prashant Doshi$^1$
\affiliations
$^1$THINC Lab, School of Computing, University of Georgia\\
\emails
\{adityas, pdoshi\}@uga.edu
}

\begin{document}

\maketitle

\begin{abstract}
This paper presents a holistic approach to attacker preference modeling from system-level audit logs using inverse reinforcement learning (IRL). Adversary modeling is an important capability in cybersecurity that lets defenders characterize behaviors of potential attackers, which enables attribution to known cyber adversary groups. Existing approaches rely on documenting an ever-evolving set of attacker tools and techniques to track known threat actors. Although attacks evolve constantly, attacker behavioral preferences are intrinsic and less volatile. Our approach learns the behavioral preferences of cyber adversaries from forensics data on their tools and techniques. We model the attacker as an expert decision-making agent with unknown behavioral preferences situated in a computer host. We leverage attack provenance graphs of audit logs to derive a state-action trajectory of the attack. We test our approach on open datasets of audit logs containing real attack data. Our results demonstrate for the first time that low-level forensics data can automatically reveal an adversary's subjective preferences, which serves as an additional dimension to modeling and documenting cyber adversaries.  Attackers' preferences tend to be invariant despite their different tools and indicate predispositions that are inherent to the attacker. As such, these inferred preferences can potentially serve as unique behavioral signatures of attackers and improve threat attribution.

\end{abstract}

%


\section{Introduction}

Sophisticated cyber attackers are increasingly targeting large organizations and critical infrastructures. These threat actors, also known as advanced persistent threats (APT), are stealthy, resourceful, and often employ novel exploitation techniques to achieve their objectives. Documenting, analyzing, and modeling such threat actors is critical for 
improving defenses against them. 
Recently, provenance graphs of audit log data has emerged as a popular computational representation for analyzing attacks by APTs~\cite{king03:backtracking,hossain17:sleuth}. 
Provenance graphs are a causal representation of interactions between kernel-level objects such as processes, threads, and files, which facilitates analysis by connecting together objects that interact. Recent efforts have adopted AI-based techniques
to automate detecting APTs~\cite{wang22:threatrace,milajerdi19:holmes}. 

Tactical information about cyber threats is critical for their detection and timely response. However, the advantage of having such specific intelligence is fleeting as attacker tools and techniques evolve constantly. Approaches to modeling adversaries lack broader insights into an attacker's behavioral characteristics and preferences. At a strategic level, past work has adopted game-theoretic~\cite{ferguson19:deception_hypergame,schlenker18:deception_game_theory} and decision-theoretic frameworks~\cite{sarraute12:pomdp_pentest,shinde24:modeling} to model cyber adversaries. However, most goal and intent recognition approaches for cybersecurity focus exclusively on end-goal recognition. These efforts do not target the broader preferences that the attacker's behavior implicitly reveals. They also rely on assumptions, such as the attacker's intent is restricted to a set of previously-known candidate reward functions~\cite{mirsky19:gr_attackgraphs,shinde21:intent_recog}.

This paper presents a novel, end-to-end approach to modeling adversary preferences from raw forensics data using inverse reinforcement learning (IRL)~\cite{arora21:survey}.  We use low-level audit logs as they are often the only attack-relevant data source in a post-breach scenario and are commonly used in cybersecurity. Our methodology leverages a provenance graph representation of the system-level audit logs. We model the attacker in a host as an expert decision-making agent in the context of a Markov decision process (MDP). Then, we utilize subgraph isomorphism to map parts of the provenance graph to attacker actions grounded in the popular MITRE ATT\&CK matrix. The ATT\&CK matrix is a comprehensive catalogue of techniques and tactics used by various attackers~\cite{strom2018mitre}. In doing so, we bridge the gap between raw security log data and the symbolic action representations that are prevalent in the decision-making models applied to cybersecurity. These mappings enable us to generate trajectories of observed attacker behavior. Subsequently, we infer an attacker's behavioral preferences from these audit logs using IRL. This is a novel application of IRL to adversary modeling, a pertinent goal in modern cyber defense. We model an adversary's preferences through behavioral features such as discoverability, duration, attributability, sophistication, and impact. {\em These informative features encompass an attacker's broader preferences at a level above their tools and techniques and serve as a unique signature of an adversary.} 

Subsequently, we utilize IRL to compute the weighting of previously mentioned preference features from the trajectories. We test this pipeline on multiple open datasets~\cite{keromytis18:darpa-e3}, which contain real cyber attacks on different target hosts. \emph{Ground truth on an adversary's preferences is usually not available! We work around this challenge by using two differing IRL techniques and analyzing inter-method agreement.} Our results demonstrate the benefit of this novel approach in extracting broader and likely invariant insights about attacker behavior from low-level log data. This automated approach to recognizing attacker preferences using IRL enables the study of behavioral aspects of cyber attackers without assumptions about their goals.

\section{Background}
\label{sec:bg}
Our approach leverages attack provenance graphs to extract trajectories of attacker behavior for inferring their preferences using IRL. In this section, we provide a brief background on provenance graphs and IRL.

\subsection{Attack Provenance Graphs}
\label{subsec:prov_bg}

APTs utilize sophisticated tools and techniques that are difficult to detect with traditional detection mechanisms~\cite{karantzas21:empirical}. Recently, the cybersecurity community has successfully applied data provenance to system-level log data for detecting such threats~\cite{king03:backtracking,hossain17:sleuth,milajerdi19:holmes,setayeshfar19:graalf}. Provenance graphs capture the interactions between system-level subjects such as processes and threads, and objects such as files, sockets, and pipes. The graphical representation aids attack investigations by restricting the search space to events causally connected to an indicator of compromise. 

\begin{figure*}[ht]
\centering
    \begin{subfigure}{0.22\linewidth}
    \centering
        \includegraphics[width=\linewidth]{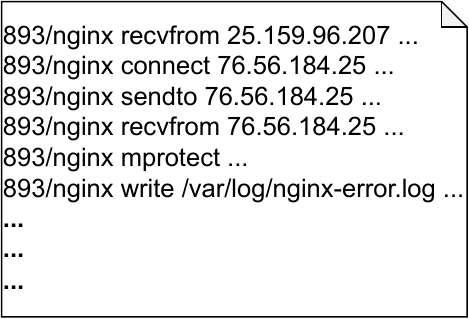}
        \caption{\small Audit logs record system-level interactions between different kernel-level objects}
        \label{subfig:log_file}
    \end{subfigure}
    \hfill
    \begin{subfigure}{0.35\linewidth}
    \centering
        \includegraphics[width=.95\linewidth]{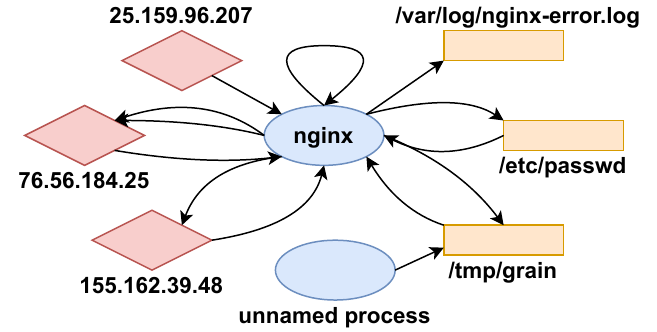}
        \caption{\small A provenance graph establishes causal relationships between related entities that interact. Here, {\em nginx} is a Web server process}
        \label{subfig:prov_graph}    
    \end{subfigure}
    \hfill
    \begin{subfigure}{0.35\linewidth}
    \centering
        \includegraphics[width=.95\linewidth]{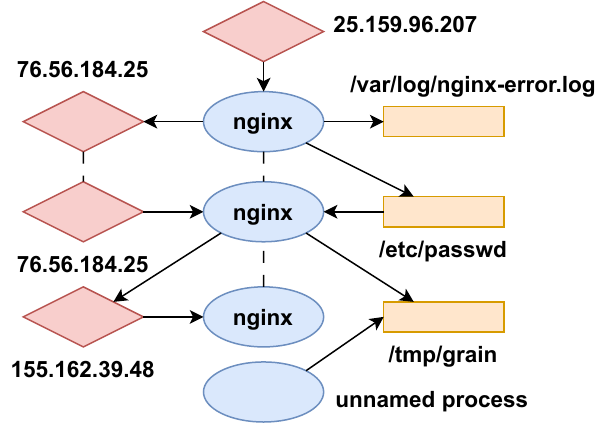}
        \caption{\small State-based versioning techniques eliminate redundant edges between entities}
        \label{subfig:vprov_graph}    
    \end{subfigure}
    \caption{\small Provenance graphs aid forensic investigations by causally connecting related entities even when they are temporally distant in logs.}
    \label{fig:prov_example}
    \vspace{-0.1in}
\end{figure*}

Formally, a provenance graph $G$ is defined as $G = \langle N, E \rangle$ where $N$ is the set of nodes representing abovementioned subjects and objects. $E = \{ e_1, e_2, ..., e_t \}$ is the set of {\em time-stamped edges} representing interactions and the direction of information flow between the nodes.
Provenance graphs facilitate impact analysis by forward-tracing the paths causally linked to a suspicious node. Starting from a node $n$ at time $t'$, the set of nodes impacted by $n$ is given by traversing all edges $e_{t \ge t'} \in E$. Similarly, backward search enables root-cause analysis that determines the source of an attack~\cite{king03:backtracking,lee13:beep}. The set of nodes that can likely influence a node $n$ at time $t'$ is similarly generated by traversing all edges $e_{t \le t'} \in E$. 
In practice, audit logging systems can produce gigabytes of log data per day~\cite{xu16:datareduction}. Several compression techniques have been proposed to address this challenge~\cite{xu16:datareduction,lee13:loggc,tang18:nodemerge}. 
Figure~\ref{fig:prov_example} illustrates this pipeline of generating a provenance graph from audit logs.

In our work, we generate an attack scenario graph starting from the source of the intrusion using forward analysis and a well-known state-based node versioning technique~\cite{hossain18:dependence}. Subsequently, we map parts of the attack scenario graph to the MITRE ATT\&CK matrix using subgraph isomorphism to generate trajectories of attacker behavior.

\subsection{Inverse RL}
\label{subsec:irl}

Inverse reinforcement learning (IRL) enables an observer to infer an expert decision-making agent's reward function from observed behavior and, optionally, the agent's policy~\cite{ng00:irl,arora21:survey}. IRL methods commonly formulate the decision making of the expert agent as a Markov decision process (MDP). The expert's MDP is defined formally as a tuple $\langle S, A, T, \gamma, R \rangle$, where $S$ is the set of states, $A$ the set of actions of the expert, $T: S \times A \times S \xrightarrow{} [0, 1]$ is the transition function, $R: S \times A \xrightarrow{} \mathbb{R}$ is the expert's unknown reward function, and $\gamma \in (0,1)$ is the discount factor. The expert agent's behavior data is available as a set of $M$ state-action trajectories of length $\mathcal{T}$, $\mathcal{X} = \{ (s_1, a_1), (s_2, a_2), ..., (s_\mathcal{T}, a_\mathcal{T}) \}$ and $|\mathcal{X}| = M$. For large state spaces, the expert's reward function is approximated linearly as $R(s, a) = w_1 \phi_1(s, a) + \dots + w_k \phi_k(s, a)$, where $s \in S$, $a \in A$, and $\phi_1, \dots, \phi_k$ are bounded basis functions $\phi: S \times A \xrightarrow{} \{0, 1\}$~\cite{Abbeel04:Apprenticeship}. {\em Basis weights $w_1, \dots, w_k$ are unknown parameters to be learned.} 


Bayesian IRL~\cite{ramachandran07:birl} is a well-known framework that assumes a prior distribution over the reward functions with i.i.d reward values, $P(R) = \prod_{s \in S, a \in A} P(R(s, a))$, and computes the posterior distribution, $P(R|\mathcal{X})$, using the expert's trajectories as follows,
\begin{align}
    P(R|\mathcal{X}) = \alpha P(\mathcal{X}|R) P(R).
    \label{eqn:birl_post}
\end{align}
Here, $\alpha$ is the normalization constant, and $P(\mathcal{X}|R)$ is the likelihood function expressed as,
\begin{align*}
    P(\mathcal{X}|R) = \prod_{m=1}^{M} \prod_{t=1}^{\mathcal{T}} \frac{e^{\beta Q^{*} (s_{t}^{m}, a_{t}^{m}; R)}}{\sum_a e^{\beta Q^{*} (s_{t}^{m}, a; R)}}.
\end{align*}
As computing the partition function contained in $\alpha$ is hard, several approaches exist to compute a point estimate of $R$ distributed according to $P(R|\mathcal{X})$~\cite{arora21:survey}. In our work, we utilize a {\em maximum-a-posteriori} (MAP) estimation for the reward function within the Bayesian IRL framework~\cite{choi11:mapbirl}. The MAP-BIRL approach computes a reward, $R_{MAP}$, that maximizes the log of the posterior in Eq.~\ref{eqn:birl_post} using a gradient method:
\begin{align*}
    R_{MAP} & = \argmax_{R} \log P(R|\mathcal{X}) \\
    & = \argmax_{R} \log P(\mathcal{X}|R) + \log P(R)
\end{align*}
Value functions $V^{*}(R)$ and $Q^{*}(R)$ are convex and differentiable almost everywhere~\cite{choi11:mapbirl}. These properties enable efficient computation of $R_{MAP}$ using a gradient method with the update rule, $R_{new} \xleftarrow{} R + \delta_t \nabla_R P(R|\mathcal{X})$. 


\begin{table*}[!ht]
\renewcommand{\arraystretch}{1.2}
\begin{center}
\begin{small}
\begin{tabular}{ |l | l | l| }
    \hline
    \textbf{Action} & \textbf{States affected} & \textbf{Description}\\
    \hline
    \hline
    \textsf{InitialAccessUser} & \textsf{AttackerActive} & Attacker gets user-level access on the target host\\
    \hline
    \multirow{2}{4em}{\textsf{InitialAccessRoot}} & \textsf{AttackerActive} & \multirow{2}{20em}{Attacker establishes privileged access} \\
    & \textsf{AttackerPrivs} & \\
    \hline
    \textsf{C2} & \textsf{C2Established} & Attacker communicates with command-and-control infrastructure \\
    \hline
    \textsf{IngressToolTransfer} & \textsf{IOCGenerated} & Attacker downloads a payload on the target host \\
    \hline
    \textsf{PrivEsc} & \textsf{AttackerPrivs} & Attacker achieves elevated privileges \\
    \hline
    \textsf{DataExfil} & -- & Attacker exfilterates sensitive data from the target \\
    \hline
    \textsf{DefenseEvasion} & \textsf{IOCGenerated} & Attacker deletes artifacts and evidence \\
    \hline
    \textsf{Exit} & \textsf{AttackerActive} & Attacker concludes the attack \\
    \hline
\end{tabular}
\caption{We model the set of attacker actions based on tactics and techniques in the MITRE ATT\&CK matrix.}
\label{tab:action_set}
\end{small}
\end{center}
\vspace{-0.1in}
\end{table*}

\section{Host-Level Cyber Threat Domain}
\label{sec:mdp}

Section~\ref{sec:bg} reviewed MAP inference in Bayesian IRL for computing the expert agent's reward function. In our work, the expert is the attacker, and the audit logs generated by the attacker's actions are the behavior data that we utilize to infer the attacker's preferences $w_i$ for reward features $\phi_i$. We present the MDP that we use to model the attacker agent next.

\subsection{Attacker Model}
\label{subsec:att_mdp}

We model the attacker's decision-making problem as an $MDP_{/R} = \langle S, A, T \rangle$, where $S$ is the set of states of the target host system, $A$ is the set of actions available to the attacker, and $T$ is the transition function representing the effect of the attacker's actions on the target host. The reward function $R$ capturing the attacker's behavior preferences is unknown to the defender and must be learned.  

Elements of the state space represent the attacker's perspective of the state of the target host. The attacker utilizes her actions to manipulate these states based on her preferences. In our model of the attacker's MDP, we define the state space using 4 state features, $S = \{ \textsf{AttackerActive, AttackerPrivs, IOCGenerated,}\\ \textsf{C2Established} \}$. 
The \textsf{AttackerActive} state feature indicates whether the attacker has established a presence on the target host. The initial value of \textsf{AttackerActive} is \texttt{false}. An attacker may utilize numerous techniques ranging from sophisticated exploits to phishing attacks to get initial access to the target host.
We model these techniques using the actions \textsf{InitialAccessUser} and \textsf{InitialAccessRoot}.
Upon the attacker's initial access, \textsf{AttackerActive} transitions to \texttt{true}. Once inside the target host, an attacker may use the \textsf{C2} action to reach the command and control (C2) infrastructure. The \textsf{C2Established} state feature represents whether the attacker has achieved communication with their C2 infrastructure. We represent the privilege level of the attacker on the target host using the \textsf{AttackerPrivs} feature. An attacker may have user-level privileges indicated by the feature value \texttt{user}, or root privileges indicated by \texttt{root}. A sophisticated attacker may achieve root privileges simultaneously while establishing initial access using the \textsf{InitialAccessRoot} action. Attackers may also utilize various privilege escalation techniques after establishing initial access. We model these techniques with the \textsf{PrivEsc} action. We model a noisy transition of \textsf{AttackerPrivs} due to \textsf{PrivEsc} to account for failed escalation attempts. Throughout the attack, the tools employed by the attacker may generate artifacts like log files. An attacker may also download additional tools such as malware stages using the \textsf{IngressToolTransfer} action. Such artifacts, also known as indicators-of-compromise, are subsequently utilized by forensics analysts to reconstruct and investigate the attack. The \textsf{IOCGenerated} state feature indicates whether the attacker's actions generate such indicators. The attacker may also erase indicators of the attack to prevent detection by end-point defense software using the \textsf{DefenseEvasion} action. Table~\ref{tab:action_set} summarizes the complete set of attacker actions that we model.


\subsection{Reward Features}
\label{subsec:rew_features}

Recall that we approximate the attacker's reward function using a linearly weighted sum of bounded basis functions, $R(s, a) = \sum_{i=1}^{k} w_i \phi_i(s, a)$. The basis functions, $\phi_i(s, a)$, are the features used to characterize the attacker's behavior. Specifically, the preference features that we model are--
\begin{enumerate}[leftmargin=*,topsep=0in,itemsep=0in]
    \item \emph{Discoverability} is the attacker's use of actions that may produce digital artifacts or other evidence. If an attacker downloads tools or malware stages to the compromised host, the attack can be discovered in an investigation. 
    \item \emph{Attributability} is the attacker's preference or lack thereof for actions that facilitate the identification of the threat actor group associated with the attack.
    \item We model \emph{sophistication} as the attacker's preference for using advanced tools and techniques. For instance, an attacker's ability to exploit privileged processes remotely is indicative of sophistication.
    \item \emph{Impact} is the attacker's preference for states and actions that can potentially cause severe consequences for the compromised host. 
    \item \emph{Evasion} is the attacker's preference for erasing digital artifacts and indicators to avoid detection.
    \item \emph{Duration} is the attacker's preference for staying active in the compromised system. 
\end{enumerate}
Table~\ref{tab:rew_features} summarizes the reward features and their associated state-action values. Though not exhaustive, the features offer deep insights into the attacker's behavior and can be inferred. 

\begin{table}[ht]
\renewcommand{\arraystretch}{1.1}
\begin{center}
\begin{small}
\begin{tabular}{ |l | l | l| }
    \hline
    \textbf{Feature name} & \textbf{States} & \textbf{Actions}\\
    \hline
    \hline
    Discoverability & Att.Active = true & IngressToolTransfer \\
    \hline
    Attributability & Att.Active = true & C2 \\
    \hline
    Sophistication & Att.Active = false & InitialAccessRoot \\
    \hline
    Impact & Att.Active = true & PrivEsc \\
    \hline
    Duration & Att.Active = true & DataExfil \\
    \hline
    \multirow{2}{4em}{Evasion} & Att.Active = true & \multirow{2}{4em}{DefenseEvasion} \\
    & IOCGen. = true & \\
    \hline
\end{tabular}
\caption{We use the reward features $\phi_i$ to model an attacker's behavioral tendencies. Here, * denotes any action.}
\label{tab:rew_features}
\end{small}
\end{center}
\vspace{-0.1in}
\end{table}

We utilize the attacker's state-action trajectory to compute the preference weights, $w_i$ of features $\phi_i$ using IRL. The values of these weights inform us about the preference ordering of each feature of the attacker's reward function. For instance, an attacker who prefers to stay undetected may prioritize deleting evidence of the attack and minimizing their attack duration over the level of impact. A likely preference ordering for such an attacker would be: $\emph{evasion} \succ \emph{sophistication} \succ \emph{impact} \succ  \emph{duration} \succ \emph{attributability} \succeq \emph{discoverability}$. In contrast, an attacker that aims to cause destruction, but is indifferent towards getting detected would prioritize maximizing impact on the compromised host resulting in a preference ordering: $\emph{impact} \succ \emph{duration} \succ \emph{sophistication} \succ \emph{attributability} \succeq \emph{discoverability} \succeq \emph{evasion}$. Such information about the attacker's implicit preferences is not directly accessible using prevailing log analysis and attack reconstruction techniques on audit log data. By modeling the contextual features of the attacker's intent, IRL enables discerning these underlying behavioral characteristics from log data.

\begin{figure*}[ht]
\centering
    \begin{subfigure}{0.24\linewidth}
    \centering
        \includegraphics[width=0.95\linewidth]{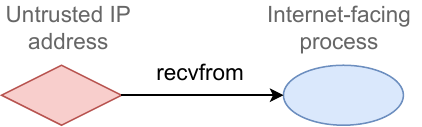}
        \caption{\small Template for the \emph{Initial Access} tactic shows the initial intrusion from a network socket to a process}
    \end{subfigure}
    \hfill
    \begin{subfigure}{0.24\linewidth}
    \centering
        \includegraphics[width=\linewidth]{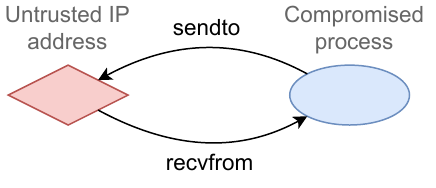}
        \caption{\small Template graph for \emph{Command and Control} shows the attacker-controlled process communicating with an external address}
    \end{subfigure}
    \hfill
    \begin{subfigure}{0.24\linewidth}
    \centering
        \includegraphics[width=0.9\linewidth]{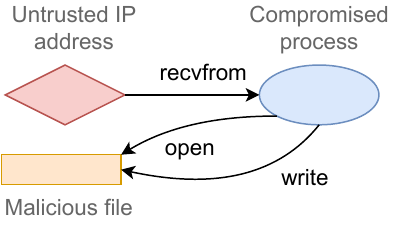}
        \caption{\small Template \emph{Ingress Tool Transfer} shows the attacker-controlled process creating an untrusted file}
    \end{subfigure}
    \hfill
    \begin{subfigure}{0.24\linewidth}
    \centering
        \includegraphics[width=0.85\linewidth]{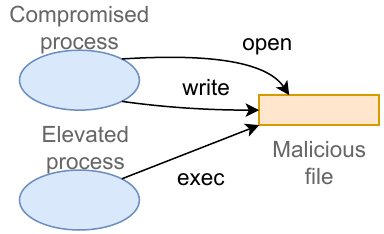}
        \caption{\small Template graph for \emph{Privilege Escalation} shows a privileged process executing an untrusted file}
    \end{subfigure}
    \caption{\small The template graphs represent tactics and techniques in the ATT\&CK matrix. We utilize these templates to identify attacker actions in the provenance graph using subgraph isomorphism.}
    \label{fig:temp_graphs}
    \vspace{-0.1in}
\end{figure*}

\section{Trajectories from Provenance Graphs}
\label{sec:sa_traj}

In cyberattack scenarios, a breach is usually detected long after the initial intrusion. Consequently, post-attack investigations must rely on log data from compromised systems to analyze the attack. In our work, we first construct provenance graphs from the log data.
We then employ subgraph isomorphism to obtain the actions defined in the attacker's MDP model for generating state-action trajectories of the attack trace.

\subsection{Obtaining Scenario Graphs from Logs}

In post-attack analyses, preliminary forensics investigations often yield information regarding the time of the initial breach, and notable artifacts such as malicious IP addresses and files. We start by processing audit log data recorded during the attack time interval. We then construct a versioned provenance graph from audit events during the attack time window using the state-versioning technique referenced in Section~\ref{subsec:prov_bg}. Notably, the state-versioning algorithm eliminates redundant events while preserving dependency between subjects and objects. We employ forward and backward search techniques on the versioned provenance graph to extract a scenario graph. 

\begin{defn}[Attack scenario]
    An attack scenario graph $G_s = \langle N_s, E_s \rangle$ is a subgraph of the provenance graph $G$ containing nodes $N_s$ and edges $E_s$ causally dependant on known attack-related nodes. It begins with first intrusion and terminates at attacker egress. 
\end{defn}

To construct $G_s$, we traverse $G$ backward from a known detection point-- like a file, to a source-- such as a network socket. We also collect process nodes that executed malicious files encountered during the backward tracing process. 
We then search forward from each source node to collect events and nodes causally related to that source node. We construct $G_s$ from the nodes and events traversed during forward search 

\begin{table}[ht!]
\renewcommand{\arraystretch}{1.1}
\begin{center}
\begin{small}
\begin{tabular}{ |l | l| }
    \hline
    \textbf{Event} & \textbf{Propagation rule} \\
    \hline
    \hline
    $S \xleftarrow{RECV} N$ & $\textit{isUntrusted}(N) \xrightarrow{} \text{tag}(S) $ \\
    $S \xrightarrow{WRITE} F$ & $\textit{isUntrusted}(S) \xrightarrow{} \text{tag}(F) $ \\
    $S \xrightarrow{EXEC} F$ & $\textit{isUntrusted}(F) \xrightarrow{} \text{tag}(S) $ \\
    $S_1 \xrightarrow{FORK} S_2$ & $\textit{isUntrusted}(S_1) \xrightarrow{} \text{tag}(S_2) $ \\
    \hline
\end{tabular}
\caption{\small Propagation rules tag \emph{untrusted} nodes based on their interactions with other nodes. (S = Subject, F = FileObject, N = NetflowObject)}
\label{tab:tag_rules}
\end{small}
\end{center}
\vspace{-0.1in}
\end{table}

The scenario graph constructed from forward and backward search may still contain subgraphs of benign events causally connected through false dependencies. To avoid false positives while mapping these attack subgraphs to attacker actions, we include an additional preprocessing step that employs tag propagation~\cite{hossain17:sleuth}. We first tag the 
nodes in the source set as \emph{untrusted}. Then, traverse the attack scenario graph and utilize the following rules to propagate the tag:
\begin{enumerate}[leftmargin=*,itemsep=0in,topsep=0in]
    \item If a subject node reads data from an untrusted network socket object, the subject node is tagged as untrusted.
    \item If an untrusted subject node writes to a file object, the file object node is tagged as untrusted.
    \item If a subject process or thread executes an untrusted file, the subject node is marked as untrusted.
    \item If a subject node is untrusted, all its child processes are tagged as untrusted.
\end{enumerate}
Table~\ref{tab:tag_rules} summarizes the events and applicable tag-propagation rules that we utilize to taint untrusted nodes. If a node is tainted, we also taint all of its subsequent versions. 
We use this information to reduce false positives when matching subgraph templates with the scenario graph to extract attacker actions.

\subsection{Extracting Trajectories using Graph Isomorphism}
The scenario graph we generate from system-level audit data contains all events causally related to the source node of an attack. We use subgraph isomorphism to match parts of the scenario graph with templates representing the attacker's actions. We call these as \emph{template graphs} and the corresponding isomorphic subgraphs in the scenario graph as \emph{action subgraphs}. Recall from Section~\ref{subsec:prov_bg} that the attacker actions we model are grounded in the MITRE ATT\&CK matrix. {\bf Consequently, we develop template graphs for the attacker actions based on the tactics and techniques in the ATT\&CK matrix.} Specifically, we build template graphs for the \emph{Initial Access}, \emph{Command and Control}, \emph{Execution}, \emph{Privilege Escalation}, and \emph{Defense Evasion} tactics. Figure~\ref{fig:temp_graphs} illustrates the template graphs for some notable actions we model. Notice that most of the attacker's actions span multiple causally related events. 
In a raw log file, long durations of unrelated activity may separate discrete events representing the same action. 
A provenance graph facilitates the direct matching of related events using subgraph isomorphism and {\em unification}, regardless of their temporal separation in a log file. Unification facilitates the matching of template and scenario subgraphs by substituting node information like process privileges and taint tags from the scenario graph into the template graphs and checking for equivalence. We unify each template graph with an isomorphic attack scenario subgraph to get a series of action subgraphs representing the attacker's action sequence.

Feature information from the nodes of the action subgraphs also facilitates the tracking of state values of the attacker's MDP. Information about a tainted subject's privileges indicates the attacker's privilege level, which determines the value of the \textsf{AttackerPrivs} state feature. Similarly, we monitor the filename information from the action subgraphs representing \emph{Ingress Tool Transfer} to track the indicators of compromise that this action generates. We utilize this information to determine the value of the \textsf{IOCGenerated} state feature. C2 IP addresses are collected similarly from the \textsf{C2} action subgraphs. Subsequently, we retrieve information about the attacker's actions and the MDP state transitions from the action subgraphs.

\subsection{Learning Attacker Preferences}

We utilize the actions and state information obtained using subgraph isomorphism to define a state-action trajectory of the attacker's observed behavior. An example sequence of an attacker's actions from such a trajectory would be, \{\emph{InitialAccessUser, C2, IngressToolTransfer, PrivEsc, ..., C2, DataExfil}\}. In this example, the attacker first achieves user-level access and then establishes C2. The attacker then downloads a payload and escalates it to root privileges. Finally, the attacker exfiltrates data from the target system. Our methodology accordingly lifts low-level system call logs to this higher level of abstraction thereby enabling the application of IRL to learn the behavioral preferences of the attacker. The host-level cyber threat domain MDP defined in Section~\ref{sec:mdp} serves as our model of the environment. The model-based MAP-BIRL reviewed in Section~\ref{subsec:irl} utilizes this MDP to infer the attacker's preferences that explain the trajectory obtained from the logs. We also use model-free MLE-IRL to learn the preferences and compare the agreement between the two techniques. For the abovementioned example trajectory, a likely preference ordering would be \emph{\{attributability, discoverability, impact, duration\} $\succ$ \{evasion, sophistication\}}.
We evaluate the performance of IRL at learning the attacker's preference function, using an empirical estimate of inverse learning error (ILE), $||V^{\pi_{E}} - V^{\hat{\pi}_{E}}||$, where $V^{\pi_{E}}$ is value of the attacker's observed trajectory using the learned Q function, and $V^{\hat{\pi}_{E}}$ is value of sampled trajectories from the learned policy. 
In the next section, we evaluate the performance of IRL toward learning attacker preferences directly from real log data.
\section{Experiments}

We described our model of the attacker's MDP in Section~\ref{sec:mdp}, and a general methodology for extracting state-action trajectories from a provenance graph representation of raw audit logs in Section~\ref{sec:sa_traj}. We utilize these trajectories with our model of the attacker's MDP to infer an attacker's hidden preference ordering for the reward features described in Section~\ref{subsec:rew_features}.

\subsection{Realistic Attack Datasets}

We evaluate our approach on large publicly available datasets from DARPA's transparent computing (TC) program~\cite{keromytis18:darpa-e3}. The data was collected from a series of red team engagements, of which Engagement 3 was publicly released. The attacks comprised APT simulations on hosts with various provenance capture software which DARPA intended to evaluate. 
We use the CADETS dataset~\cite{strnad19:cadets} recorded on a FreeBSD host, and the THEIA dataset~\cite{fazzini17:theia} recorded on a Linux host. We evaluate our approach on \textbf{4} provenance graphs automatically constructed from the CADETS logs and \textbf{2} from the THEIA logs. These graphs contain data from separate APT attacks. Table~\ref{tab:dataset} shows the size of each attack scenario graph with the attack storyline obtained from log analysis and after-action reports.

\begin{table}[!ht]
\renewcommand{\arraystretch}{1.1}
\begin{scriptsize}
\begin{center}
\begin{tabular}{| l | c | c | c |}
    \hline
    \textbf{Datasets} & \bm{$| N_s |$} & \bm{$| E_s |$} & \textbf{Attack storyline and Ground truth} \\
    \hline
    \hline
    CADETS-1 & 8,394 & 21,394 & $RE \xrightarrow{} C2 \xrightarrow{} PE \xrightarrow{} DE$ \\
    & & & {\bf (Features: Sop, Att, Imp, Dis)} \\
    \hline
    CADETS-2 & 1,734 & 9,267 & $UA \xrightarrow{} C2$   {\bf (Features: Att, Dis)} \\
    \hline
    CADETS-3 & 44,974 & 98,584 & $UA \xrightarrow{} C2 \xrightarrow{} PE \xrightarrow{} DE$ \\
    & & & {\bf (Features: Att, Imp, Dis, Eva)} \\
    \hline
    CADETS-4 & 196,502 & 469,094 & $UA \xrightarrow{} C2 \xrightarrow{} PE$ \\
    & & & {\bf (Features: Att, Imp, Dis)} \\
    \hline
    THEIA-1 & 824 & 911 & $UA \xrightarrow{} C2 \xrightarrow{} PE$\\
    & & & {\bf (Features: Att, Imp, Dis)} \\
    \hline
    THEIA-2 & 30,504 & 70,300 & $UA \xrightarrow{} C2 \xrightarrow{} PE \xrightarrow{} DE$ \\
    & & & {\bf (Features: Att, Imp, Dis, Eva)} \\
    \hline
\end{tabular}
\caption{\small Size of each dataset's attack scenario graph with the attack storyline (RE = Root exploit, UA = User-level access, C2 = Command and control, PE = Privilege elevation, DE = Defense evasion). Associated behavioral features are also shown (Sop = Sophistication, Att = Attributability, Imp = Impact, Dis = Discoverability, Eva = Evasion, Dur = Duration), which serves as the ground truth.}
\label{tab:dataset}
\end{center}
\end{scriptsize}
\vspace{-0.1in}
\end{table}

The DARPA TC dataset is widely used for evaluating provenance graph-based APT detection. While some inconsistencies were reported in the data caused by the target hosts occasionally crashing during process-injection attacks, the dataset is a benchmark in the cybersecurity community due to a lack of realistic attack log data. We could infer attacker preferences from the available data using our approach. We show detailed trajectories extracted from each dataset in Appendix A. 

\begin{figure*}[!ht]
\centering
    \begin{subfigure}{0.32\linewidth}
    \centering
        \includegraphics[width=0.8\linewidth]{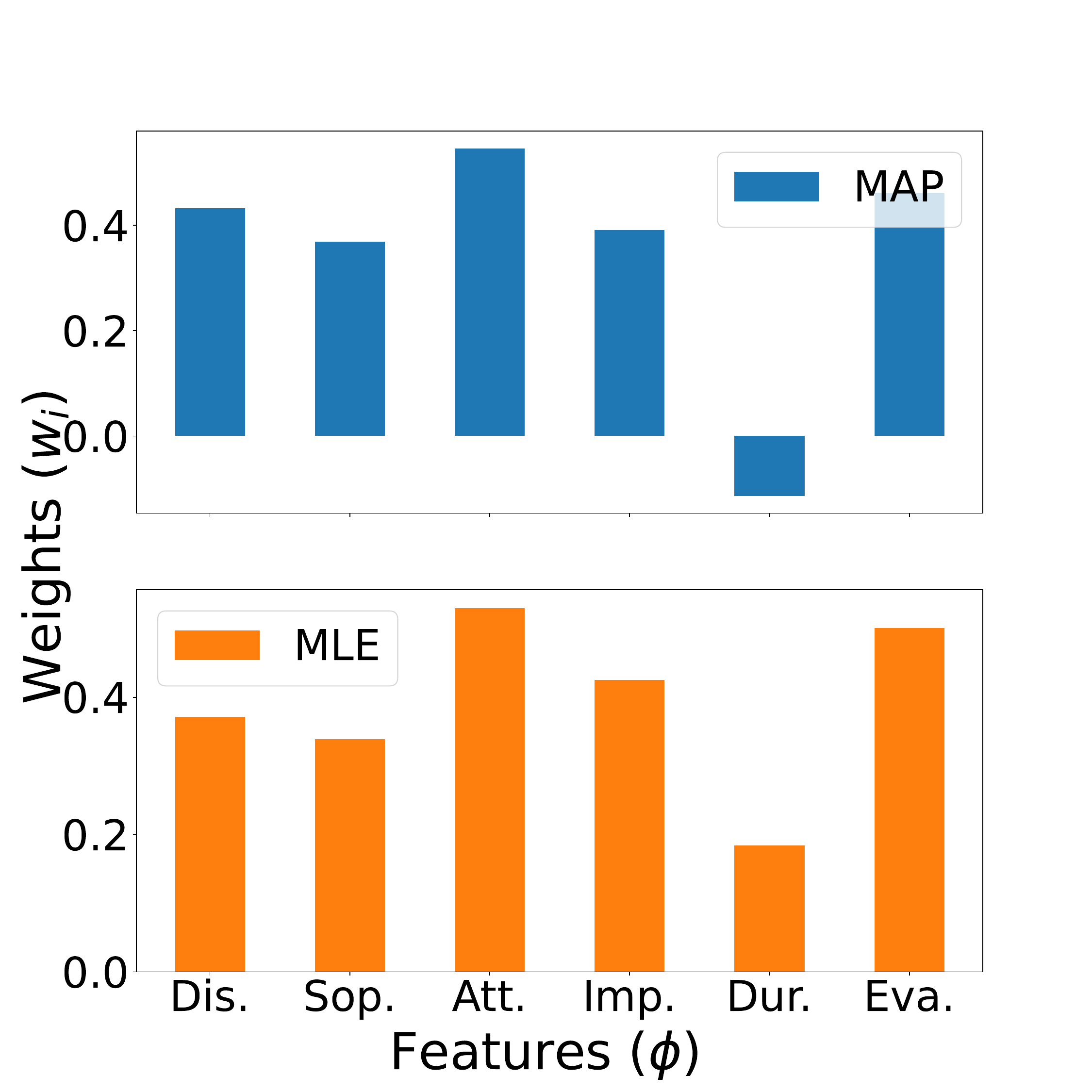}
        \caption{Behavioral signature from MAP for CADETS-1 indicates the ordering, 
        \emph{\{attributability, evasion, discoverability, impact, sophistication\} $\succ$ duration}}
        \label{subfig:cadets1_weights}
    \end{subfigure}
    \hfill
    \begin{subfigure}{0.32\linewidth}
    \centering
        \includegraphics[width=0.8\linewidth]{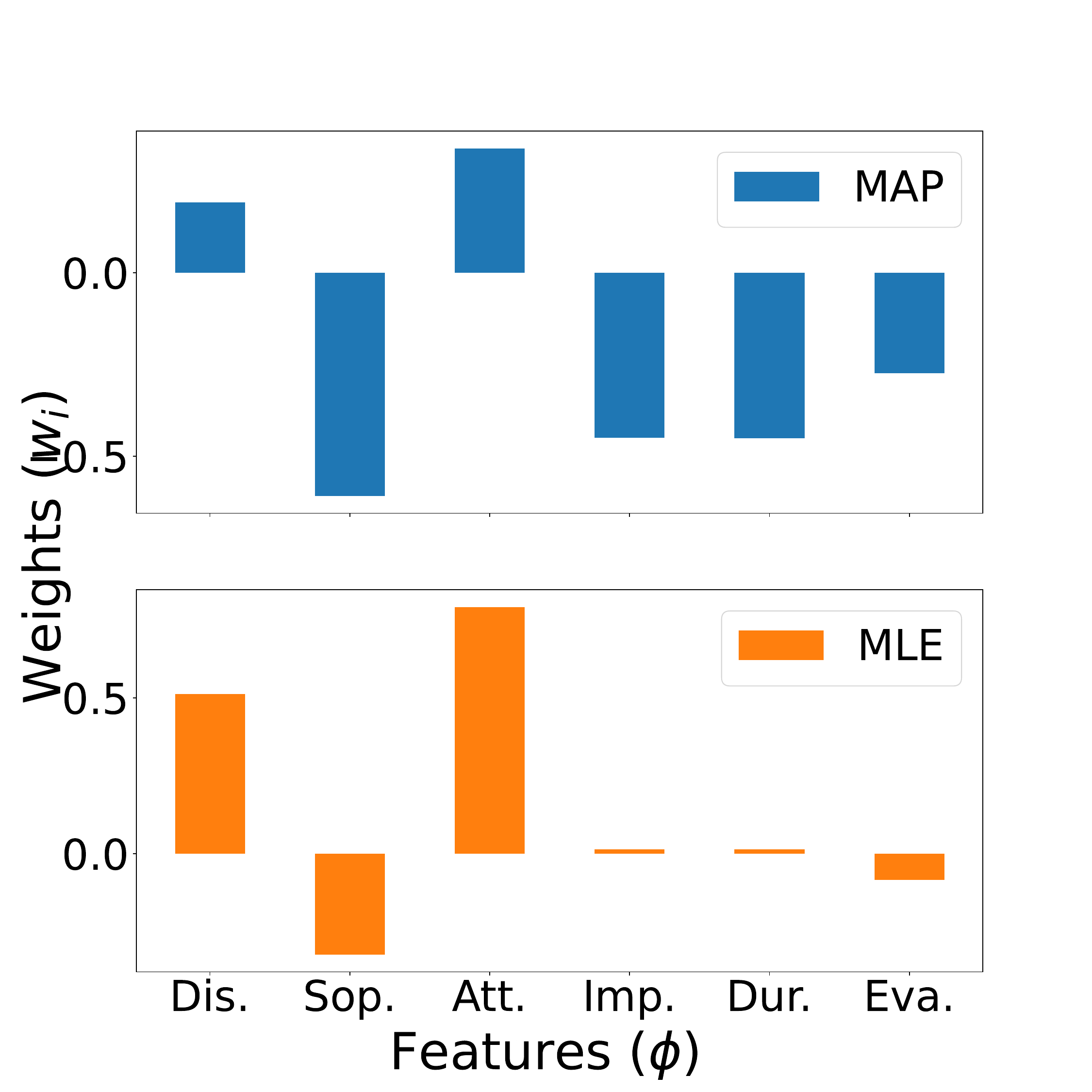}
        \caption{Behavioral signature from MAP for CADETS-2 indicates the ordering, \emph{\{attributability, discoverability\} $\succ$ \{evasion, impact, duration,  sophistication\}}}
        \label{subfig:cadets2_weights}
    \end{subfigure}
    \hfill
    \begin{subfigure}{0.32\linewidth}
    \centering
        \includegraphics[width=0.8\linewidth]{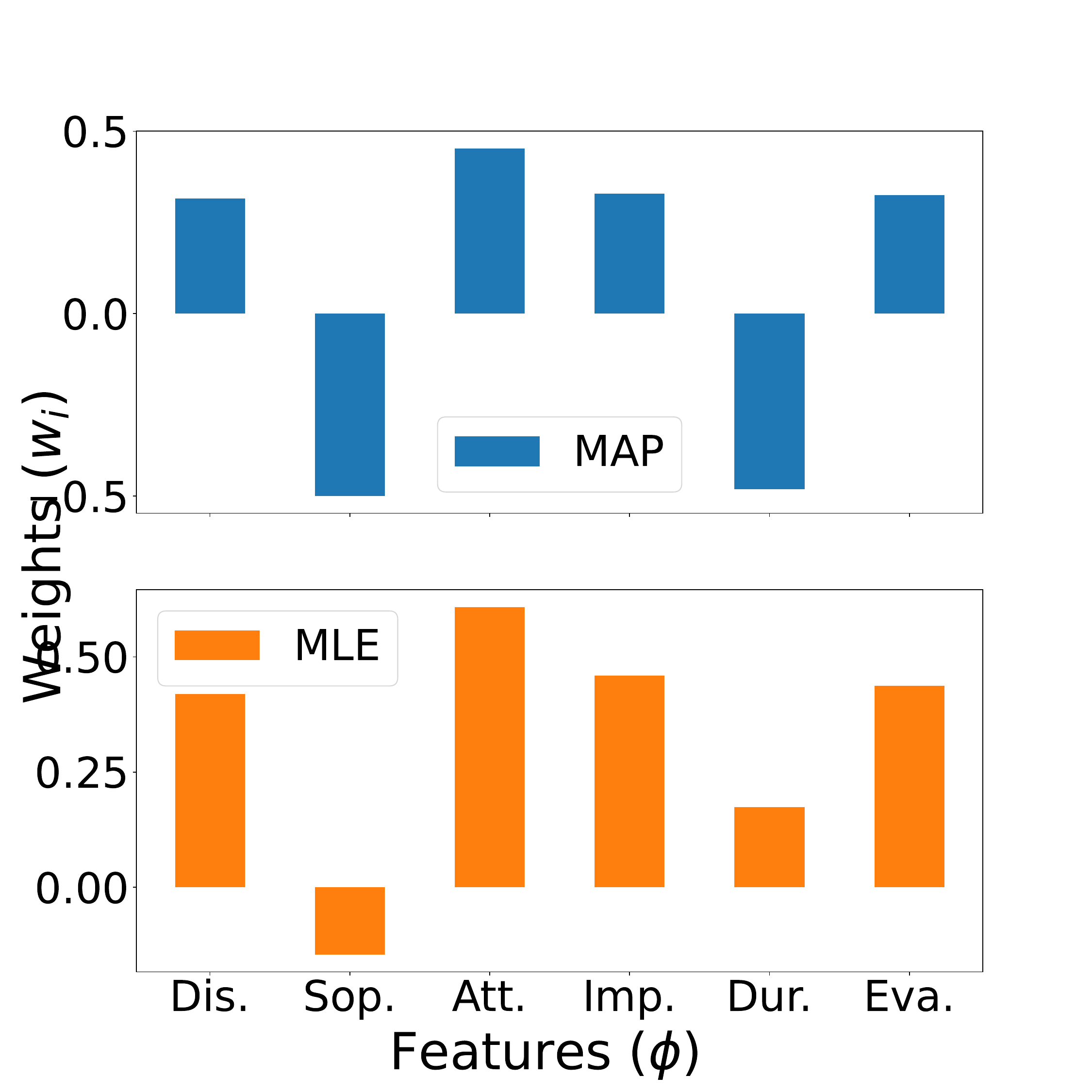}
        \caption{Behavioral signature from MAP for CADETS-3 indicates the ordering, \emph{\{attributability, impact, evasion, discoverability\} $\succ$ \{sophistication, duration\}}}
        \label{subfig:cadets3_weights}
    \end{subfigure}

    \begin{subfigure}{0.32\linewidth}
    \centering
        \includegraphics[width=0.8\linewidth]{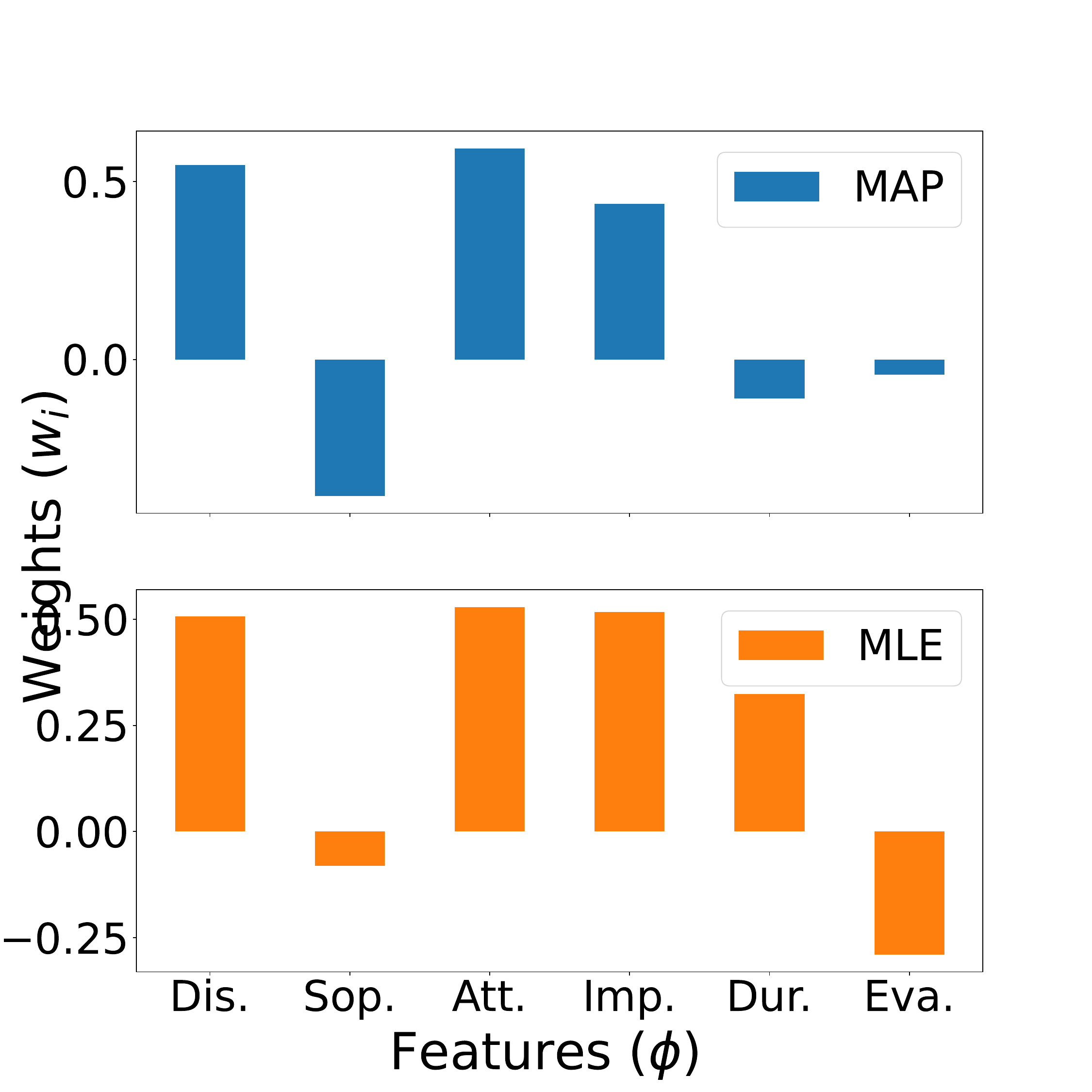}
        \caption{Behavioral signature from MAP for CADETS-4 indicates the ordering, \emph{\{attributability, discoverability, impact\} $\succ$ \{evasion, duration\} $\succ$ sophistication}}
        \label{subfig:cadets4_weights}
    \end{subfigure}
    \hfill
    \begin{subfigure}{0.32\linewidth}
    \centering
        \includegraphics[width=0.8\linewidth]{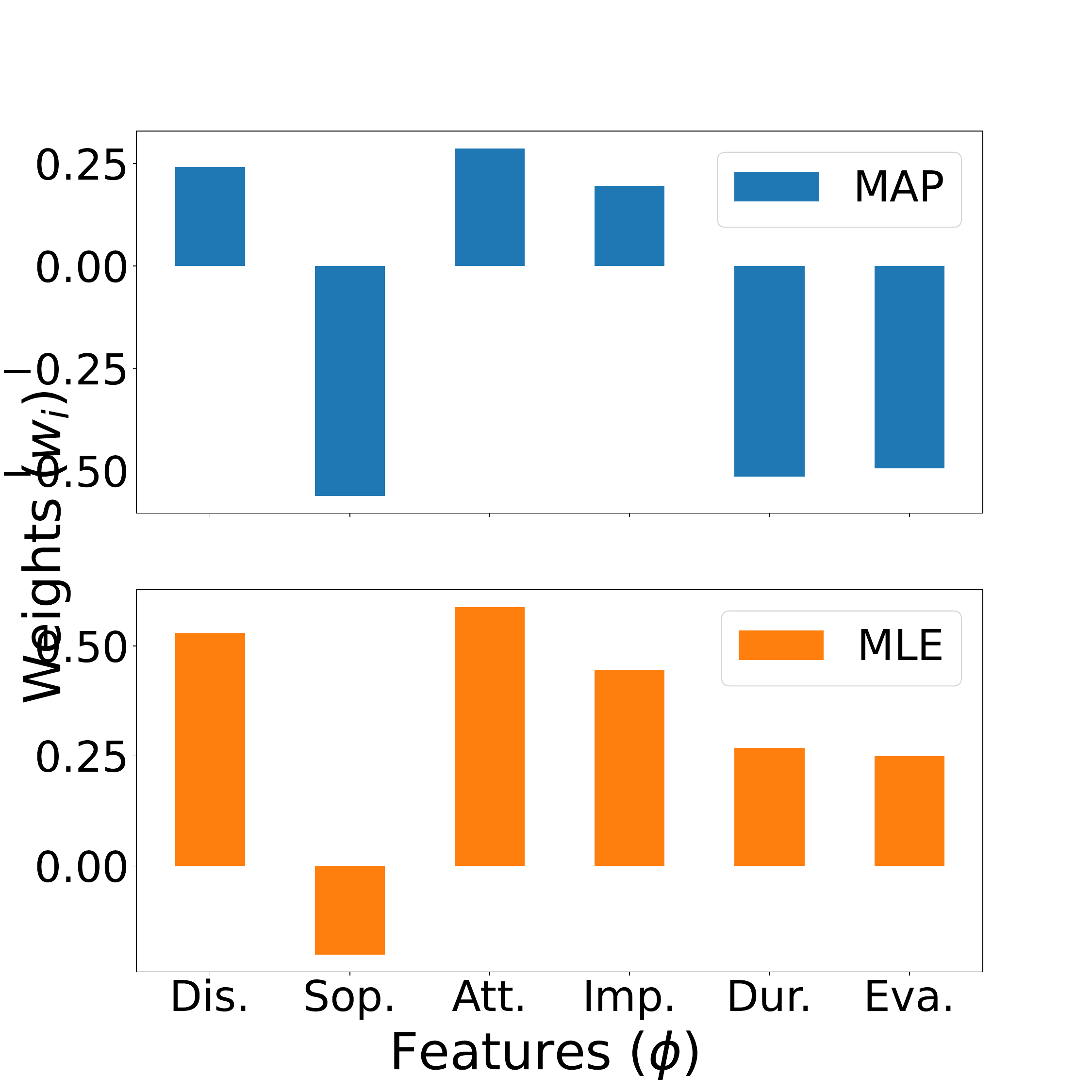}
        \caption{Behavioral signature from MAP for THEIA-1 indicates the ordering, \emph{\{attributability, discoverability, impact\} $\succ$ \{evasion, duration, sophistication\}}}
        \label{subfig:theia1_weights}
    \end{subfigure}
    \hfill
    \begin{subfigure}{0.32\linewidth}
    \centering
        \includegraphics[width=0.8\linewidth]{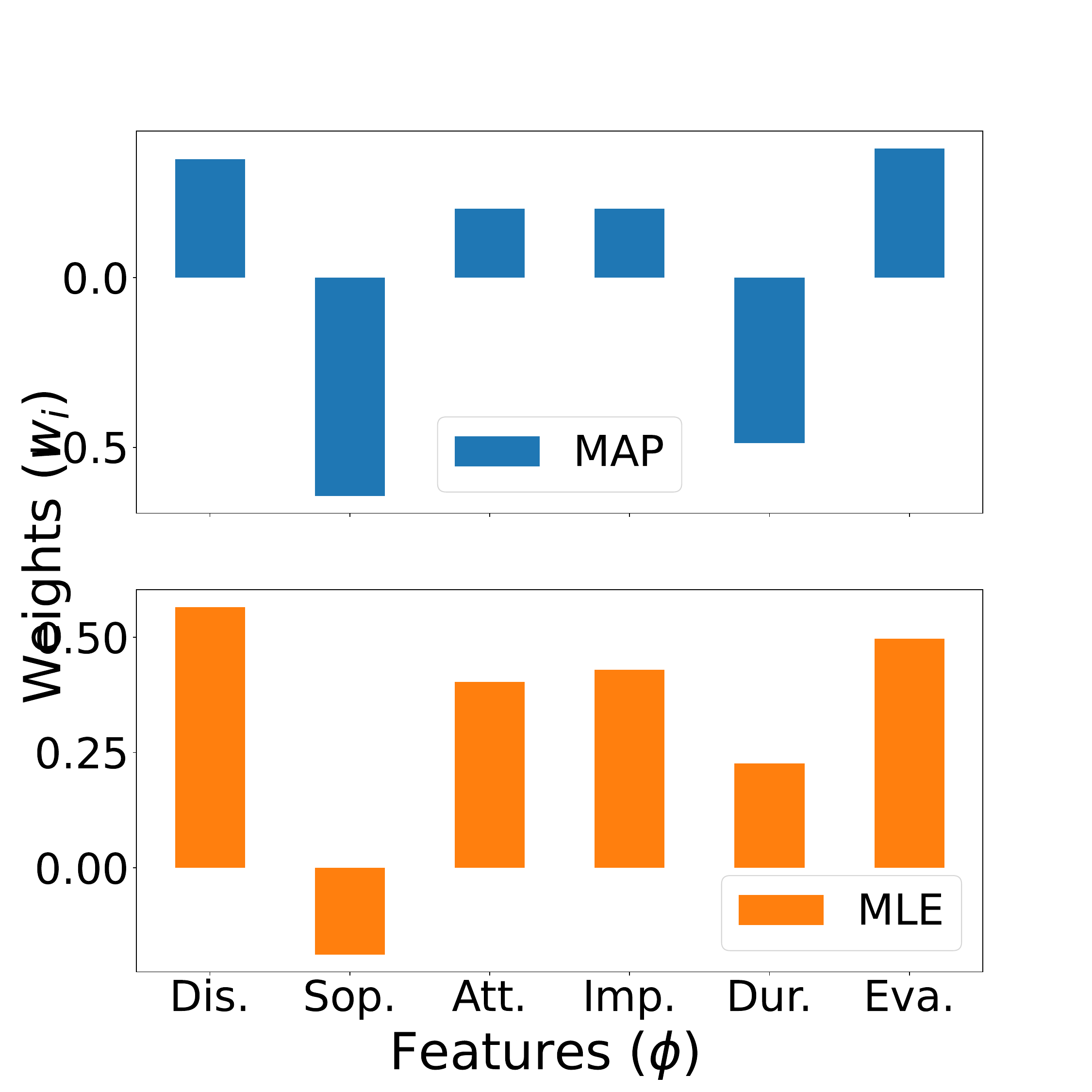}
        \caption{Behavioral signature from MAP for THEIA-2 indicates the ordering, \emph{\{evasion, discoverability, attributability, impact\} $\succ$ \{duration, sophistication\}}}
        \label{subfig:theia2_weights}
    \end{subfigure}
    \caption{The reward functions inferred from the trajectories of different attackers using MAP-BIRL show their behavioral preferences.}
    \label{fig:pref_weights}
    \vspace{-0.1in}
\end{figure*}

\subsection{Learned Preferences}

We generated provenance graphs for each APT attack in the CADETS and THEIA datasets and stored them in a Neo4j database. We then extracted state-action trajectories for each attack using Cypher queries for subgraph isomorphism. To infer the attacker's preferences, we use MAP-BIRL as well as {\em model-free} maximum likelihood estimation (MLE) approach to IRL~\cite{Jain19:Model}. We then use Mean Shift clustering to group the preference features according to their learned weights. We estimate ILE using 1000 sampled trajectories. Both techniques are effective in learning the preference function as indicated by the low ILE values, with MAP-BIRL performing better on more of the datasets.

\begin{table}[ht]
\renewcommand{\arraystretch}{1.1}
\begin{scriptsize}
\begin{center}
\begin{tabular}{| l | c | c | c |}
    \hline
    \textbf{Datasets} & \textbf{Spearman's} $\rho$ & \multicolumn{2}{c|}{\bf Inverse learning error}\\  
    \cline{3-4}
    & & MAP-BIRL & MLE-IRL \\
   \hline
    CADETS-1 & 0.94, $p < 0.005$ & 1.75 $\pm$ 1.44 & 1.46 $\pm$ 1.66 \\ 
    CADETS-2 & 0.82, $p < 0.05$ & 3.74 $\pm$ 3.5 & 6.21 $\pm$ 3.45 \\  
    CADETS-3 & 0.94, $p < 0.005$ & 4.18 $\pm$ 2.5 & 9.61 $\pm$ 8.76 \\ 
    CADETS-4 & 0.77, $p < 0.08^\dagger$ & 1.2 $\pm$ 1.35 & 3.1 $\pm$ 5.44 \\ 
    THEIA-1 & 0.94, $p < 0.005$ & 4.4 $\pm$ 4.26 & 11.77 $\pm$ 7.08 \\  
    THEIA-2 & 0.94, $p < 0.005$ & 3.53 $\pm$ 3.58 & 10.71 $\pm$ 6.48 \\  
    \hline
    \end{tabular}
\caption{\small Spearman's rank correlation coefficient ($\rho$) measures the agreement between the feature weights learned by MAP-BIRL and MLE-IRL.  $\dagger$-- may not correlate. MAP-BIRL exhibits a lower ILE for all but one dataset.}
\label{tab:agreement}
\end{center}
\end{scriptsize}
\vspace{-0.1in}
\end{table}

Figure~\ref{fig:pref_weights} shows the normalized weights representing each attacker's behavioral preferences learned using both methods. Logs from CADETS-1 consisted of \textbf{18} attacker actions. The attacker deployed a remote exploit to get \emph{root}-level access via an \emph{nginx} server ({\sf InitialAccessRoot}) demonstrating high \emph{sophistication}.
The attacker's preference to escalate a malware payload, erase evidence and C2 with multiple IP addresses indicated high \emph{impact}, \emph{evasion}, and \emph{attributability} as shown in Table~\ref{tab:dataset}.
Both MAP and MLE IRL correctly infer these preferences as shown in Fig.~\ref{subfig:cadets1_weights}. In the CADETS-2 attack, our methodology identified \textbf{7} attacker actions. The attacker initially gained user-level access ({\sf InitialAccessUser}) and immediately established C2. Subsequently, the attacker downloaded a payload ({\sf IngressToolTransfer}) and concluded the attack. 
The attacker's failure to elevate privileges and erase the payload was correctly inferred by MAP via negative weights for \emph{impact} and \emph{evasion} as shown in Fig.~\ref{subfig:cadets2_weights}.
The CADETS-3 attack contained \textbf{55} actions. Similar to CADETS-2, the attacker gained access by exploiting a Web-facing application. Subsequently, the attacker downloaded multiple payloads for process injection but failed, and instead ran an elevated process. 
The attacker's preference for subsequent privilege escalation instead of an initial root-level exploit was correctly inferred by MAP-BIRL as a lack of \emph{sophistication}.
The CADETS-4 and THEIA-1 attacks were similar, consisting of \textbf{17} and \textbf{12} actions respectively. The attackers gained user-level access, established C2, and elevated privileges. However, both failed to erase their respective payloads to avoid detection. 
Both IRL techniques correctly identified this behavioral preference as indicated by lower values for \emph{evasion} in Figs.~\ref{subfig:cadets4_weights} and~\ref{subfig:theia1_weights}. Finally, the THEIA-2 attack contained \textbf{11} actions.
The attacker gained user-level access and downloaded payloads for injection. However, the attacker promptly deleted all except one payload. Consequently, Fig.~\ref{subfig:theia2_weights} shows higher preferences for \emph{discoverability} and \emph{evasion}. 
Similar to CADETS-3,4 and THEIA-1, a lack of \emph{sophistication} was also observed. None of the attacks contained data exfiltration or similar actions requiring prolonged attacker presence. Consequently, the weights for \emph{duration} were low for all attacks. 

\begin{table}[ht]
\renewcommand{\arraystretch}{1.1}
\begin{scriptsize}
\begin{center}
\begin{tabular}{| l | c | c |}
    \hline
    \textbf{Datasets} & \textbf{Ground Truth} & \textbf{Preference Ordering}\\
    & & \textbf{Learned by MAP-BIRL} \\
    \hline
    CADETS-1 & Sop, Att, Imp, Dis & \emph{\{Att, Eva, Dis, Imp, Sop\} $\succ$ Dur} \\ 
    CADETS-2 & Att, Dis & \emph{\{Att, Dis\} $\succ$ \{Eva, Imp, Dur, Sop\}} \\  
    CADETS-3 & Att, Imp, Dis, Eva & \emph{\{Att, Imp, Eva, Dis\} $\succ$ \{Sop, Dur\}} \\ 
    CADETS-4 & Att, Imp, Dis & \emph{\{Att, Dis, Imp\} $\succ$ \{Eva, Dur\} $\succ$ Sop} \\ 
    THEIA-1 & Att, Imp, Dis & \emph{\{Att, Dis, Imp\} $\succ$ \{Eva, Dur, Sop\}} \\  
    THEIA-2 & Att, Imp, Dis, Eva & \emph{\{Eva, Dis, Att, Imp\} $\succ$ \{Dur, Sop\}} \\  
    \hline
    \end{tabular}
\caption{\small The preference orderings learned by MAP-BIRL is consistent with the features emphasized in the ground truth.}
\label{tab:ordering}
\end{center}
\end{scriptsize}
\vspace{-0.1in}
\end{table}

Table~\ref{tab:agreement} shows the rank correlation between the preferences learned by both IRL techniques. The table also shows that both techniques were effective in learning the attacker's reward function as indicated by the low values of ILE. Note that the preferences inferred by our methodology are consistent with the ground-truth as shown in Table~\ref{tab:ordering}. These weights can serve as unique behavior-generating signatures of the attackers.


\section{Related Work}
Recognizing an attacker's intent from forensics data is a topic of much interest at the intersection of cyber security and AI. 

\noindent \textbf{Log Analysis:} Several recent works in cybersecurity adopt provenance graphs for APT detection~\cite{hossain17:sleuth,milajerdi19:poirot,wang22:threatrace,cheng24:kairos}. HOLMES~\cite{milajerdi19:holmes} is one such relevant approach that explains APT campaigns at a tactical level. However, these approaches aim to identify APT activity and reconstruct attacks from provenance graphs. Our work goes beyond attack detection and models APT behavior to supplement post-attack investigations with deeper insights into attacker preferences.


\noindent \textbf{AI-based Intent Recognition:}
Recently, AI-based techniques are also being applied to attacker intent recognition~\cite{kassa24:cybercrime}. One such approach proposes an AI-based methodology to identify attack phases from system call logs using an HMM and learned classifiers~\cite{abuodeh21:cyberian}. Another approach employed an HMM to recognize tactics in the MITRE ATT\&CK matrix from sensor alerts~\cite{zhang09:hmm}. Instead, we model these tactics and phases as actions and learn the intrinsic behavioral preferences of an attacker from them. Another interesting work adopts the I-POMDP$_\mathcal{X}$ framework for attacker intent recognition on a honeypot system~\cite{shinde21:intent_recog}. The I-POMDP$_\mathcal{X}$-based defender employs deception to actively infer an attacker's intent from a predefined set specified a priori. In contrast, our work does not make such assumptions about the attacker's intent and learns preferences for abstract behavioral features. As such, our approach to modeling these preferences using IRL differs significantly from conventional approaches in cybersecurity.
\section{Conclusion}

The lack of representative data on adversarial intent is a known challenge in cybersecurity. Low-level forensics logs are often the only source of attack-relevant data. Conventional approaches to understanding adversary tools and techniques typically do not provide long-term insights into adversary behavior as attacker tools and techniques evolve constantly. However, their behavioral tendencies are long-lasting and independent of their specific tools. Consequently, automated ways of extracting higher-level (deeper) insights into adversary behavior from low-level data is very valuable to modern cyber defense. Our AI-anchored methodology instruments a {\em novel} use case of IRL to model cyber adversaries. The significant results demonstrate the efficacy of the methodology and IRL's effectiveness in correctly learning adversary behavior from log data.     

Insights into an adversary's behavioral tendencies will enable defenders to orient their defenses appropriately to prevent future attacks. Toward this, future work could utilize the learned preferences to engage in forward RL on simulations of various host configurations to predict how an attack from the adversary would unfold.

\bibliographystyle{named}
\bibliography{refs}

\section{Appendix}

\begin{figure}[ht]
\centering
    \includegraphics[width=0.9\linewidth]{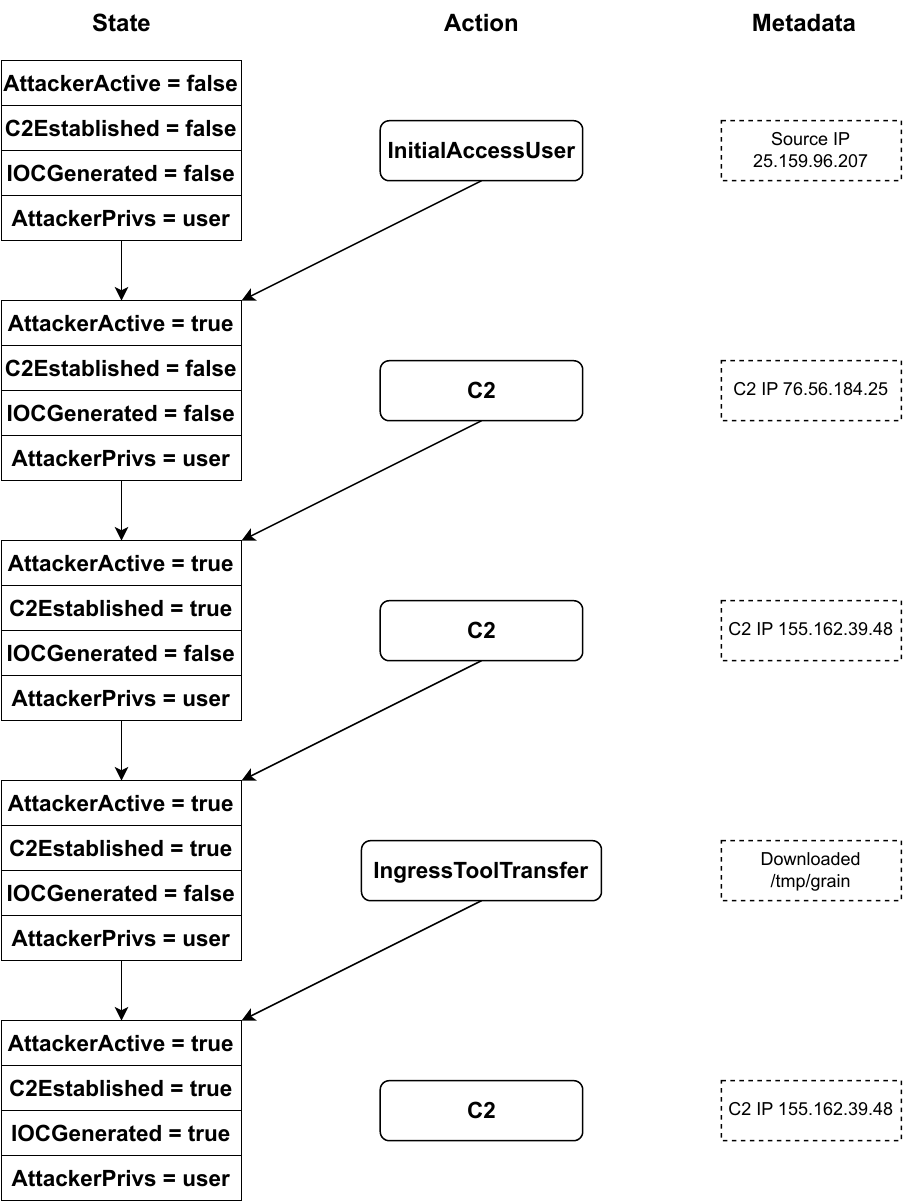}
    \caption{The state-action trajectory for the CADETS-2 attack}
    \label{fig:cadets2}
\end{figure}

\subsection{CADETS-1}
\begin{figure}[ht]
\centering
    \includegraphics[width=0.9\linewidth]{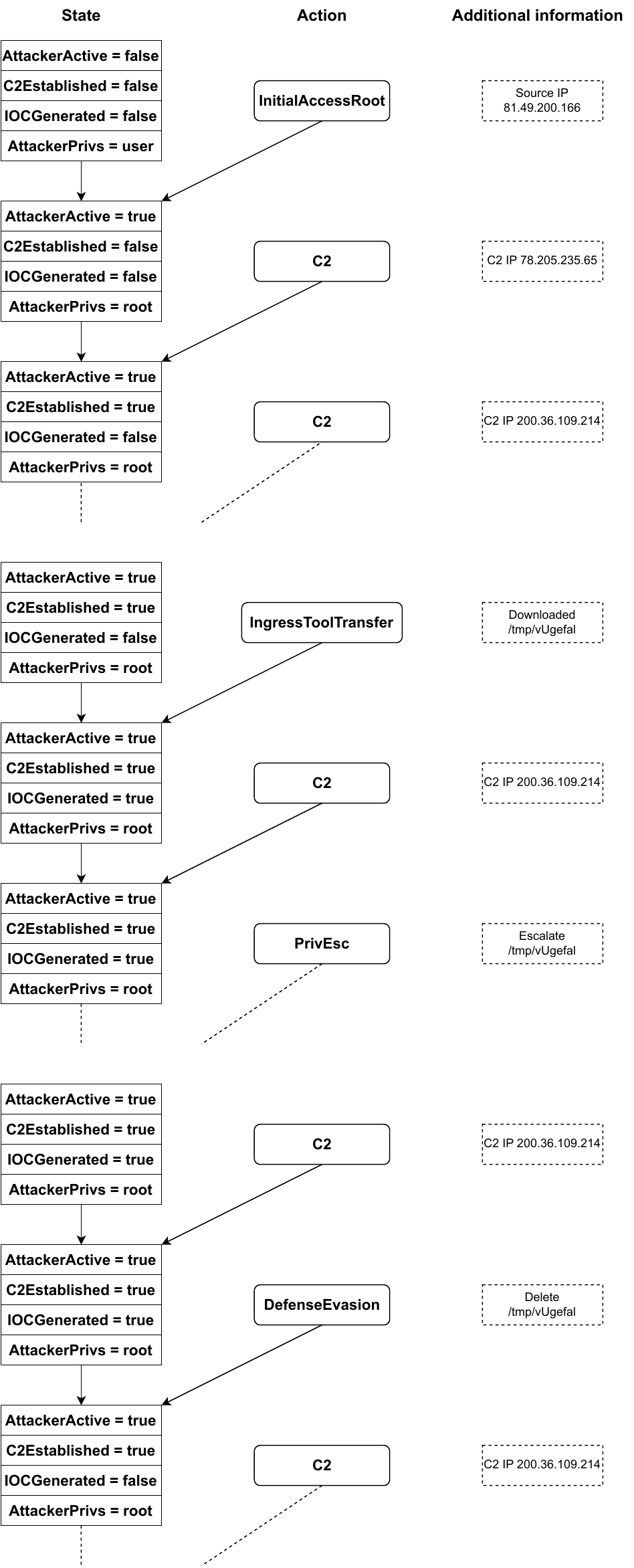}
    \caption{The state-action trajectory for the CADETS-1 attack}
    \label{fig:cadets1}
\end{figure}

The CADETS-1 attacker exploited an internet-facing application running at elevated privileges for initial access. The attacker then established command and control with the IP addresses 78.205.235.65 and 200.36.109.214. Next, the attacker downloaded the file /tmp/vUgefal. This file was an APT stage which the attacker executed with root-level privileges. The attacker then deleted /tmp/vUgefal to avoid detection. Subsequently, the attacker downloaded another APT stage /var/log/devc before exiting the system. Figure~\ref{fig:cadets1} shows the important state-action pairs for this trajectory.

\subsection{CADETS-2}

\begin{figure*}[ht]
\centering
    \includegraphics[width=0.8\linewidth]{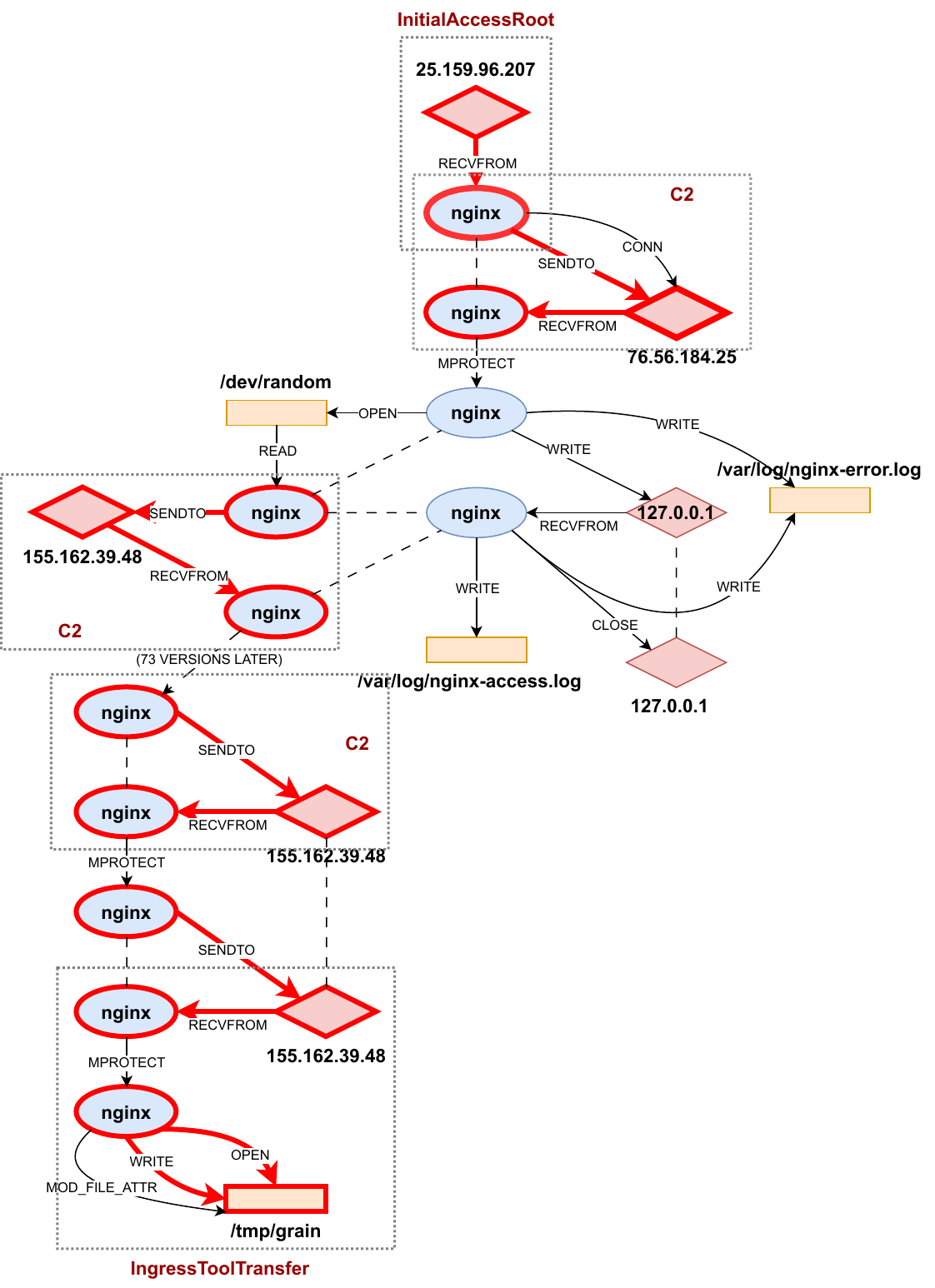}
    \caption{A small subgraph of the state-versioned provenance graph for the CADETS-2 attack shows the attacker's activity on the target system. The nodes and edges highlighted in red match the subgraph templates for attacker actions in the MDP model. The dotted boxes surrounding those subgraphs indicate the action that was identified using subgraph isomorphism}
    \label{fig:cadets2_prov}
\end{figure*}

Figure~\ref{fig:cadets2} shows the trajectory for the CADETS-2 attacker. The attacker established user-level access by exploiting an internet-facing application. Next, the attacker reached out to the IP addresses 76.56.184.25 and 155.162.39.48. Finally, the attacker downloaded the file /tmp/grain, an APT stage. Figure~\ref{fig:cadets2_prov} shows a small provenance subgraph of the CADETS-2 attack and attacker actions obtained using subgraph isomorphism with the template graphs.

\subsection{CADETS-3}

\begin{figure}[ht]
\centering
    \includegraphics[width=0.9\linewidth]{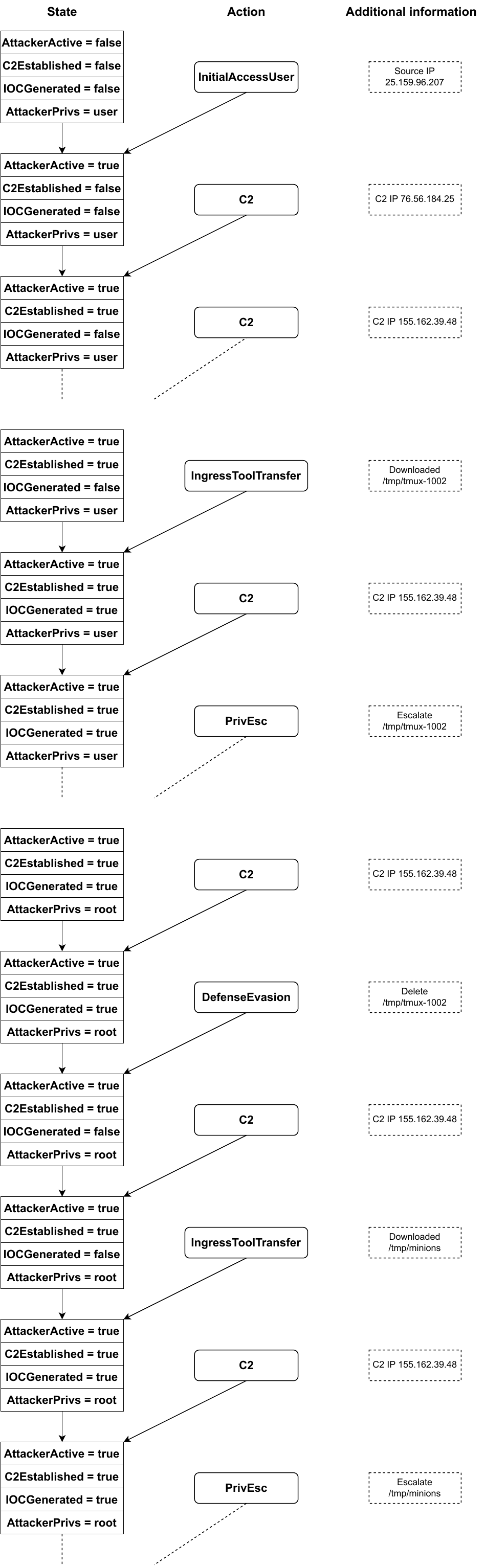}
    \caption{The state-action trajectory for the CADETS-3 attack}
    \label{fig:cadets3}
\end{figure}

The CADETS-3 attacker began with user-level access to the target. The attack lasted for 53 steps. Throughout the attack, the attacker communicated with 76.56.184.25, 155.162.39.48, 53.158.101.118, and 192.113.144.28 for C2. Additionally, the attacker downloaded multiple APT stages and attempted to escalate them. As a result, various indicators of compromise were generated. Specifically, /tmp/tmux-1002, /tmp/minions, /tmp/font, /tmp/XIM, /var/log/netlog, /var/log/sendmail, /tmp/main, and /tmp/test were the files containing APT stages that the attacker downloaded. Subsequently, the attacker deleted all files except /tmp/minions. Figure~\ref{fig:cadets3} shows the important state-action pairs for this trajectory.

\subsection{CADETS-4}

\begin{figure}[ht]
\centering
    \includegraphics[width=0.9\linewidth]{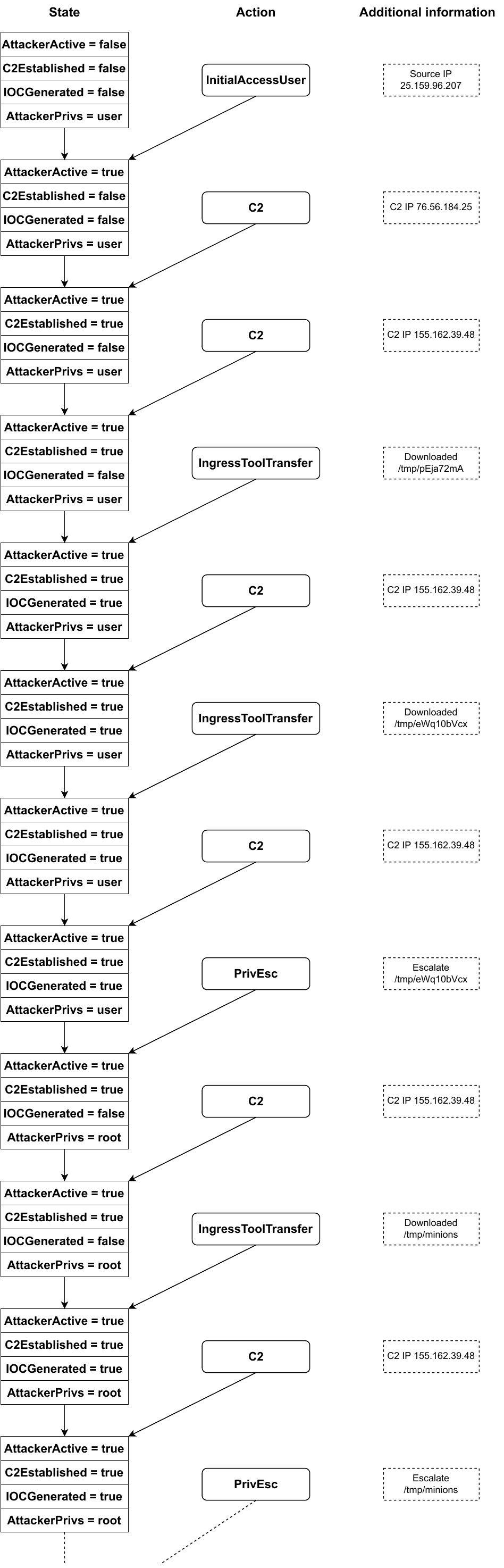}
    \caption{The state-action trajectory for the CADETS-4 attack}
    \label{fig:cadets4}
\end{figure}

The CADETS-4 attacker started the attack with user-level access to the target. The attacker established command and control with the IP addresses 76.56.184.25, 155.162.39.48, and 53.158.101.118. The attacker also downloaded the APT stages /tmp/pEja72mA, /tmp/eWq10bVcx, /tmp/memhelp.so, /tmp/eraseme, and /tmp/done.so. Only /tmp/pEja72mA was escalated to root privileges. The attacker also did not attempt to erase any downloaded APT stages. Figure~\ref{fig:cadets3} shows the notable state-action pairs for this trajectory.

\subsection{THEIA-1}

\begin{figure}[ht]
\centering
    \includegraphics[width=0.9\linewidth]{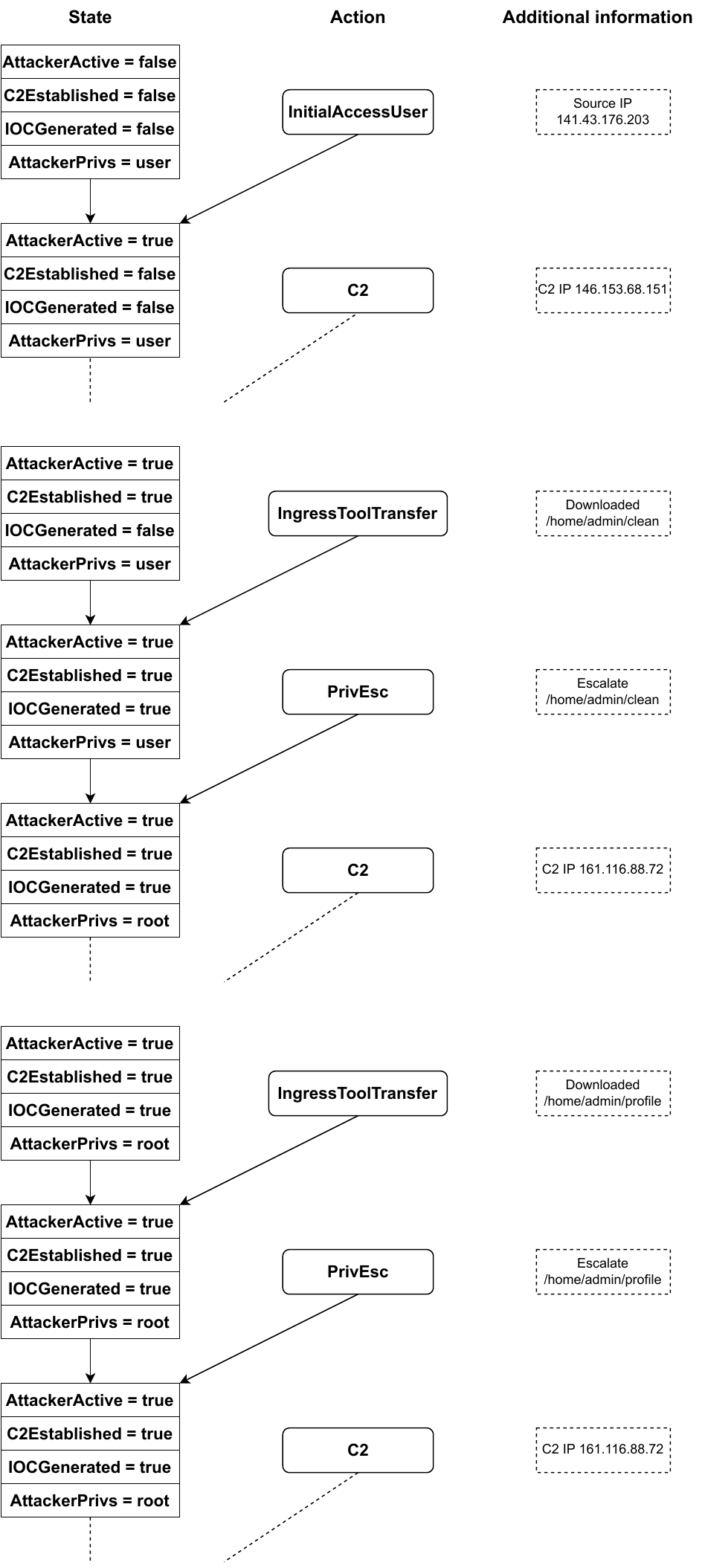}
    \caption{The state-action trajectory for the THEIA-1 attack}
    \label{fig:theia1}
\end{figure}

The THEIA-1 attacker also started the attack by exploiting a user-level application on the target. The attacker used the IP addresses 146.153.68.151 and 161.116.88.72 for C2. The attacker then downloaded the APT stages /home/admin/clean and /home/admin/profile. Both APT stages were elevated to root privileges for execution. Similar to CADETS-4, the attacker did not attempt to avoid detection by deleting the APT stages once they were executed. Figure~\ref{fig:theia1} shows the THEIA-1 trajectory.

\subsection{THEIA-2}

\begin{figure}[ht]
\centering
    \includegraphics[width=0.9\linewidth]{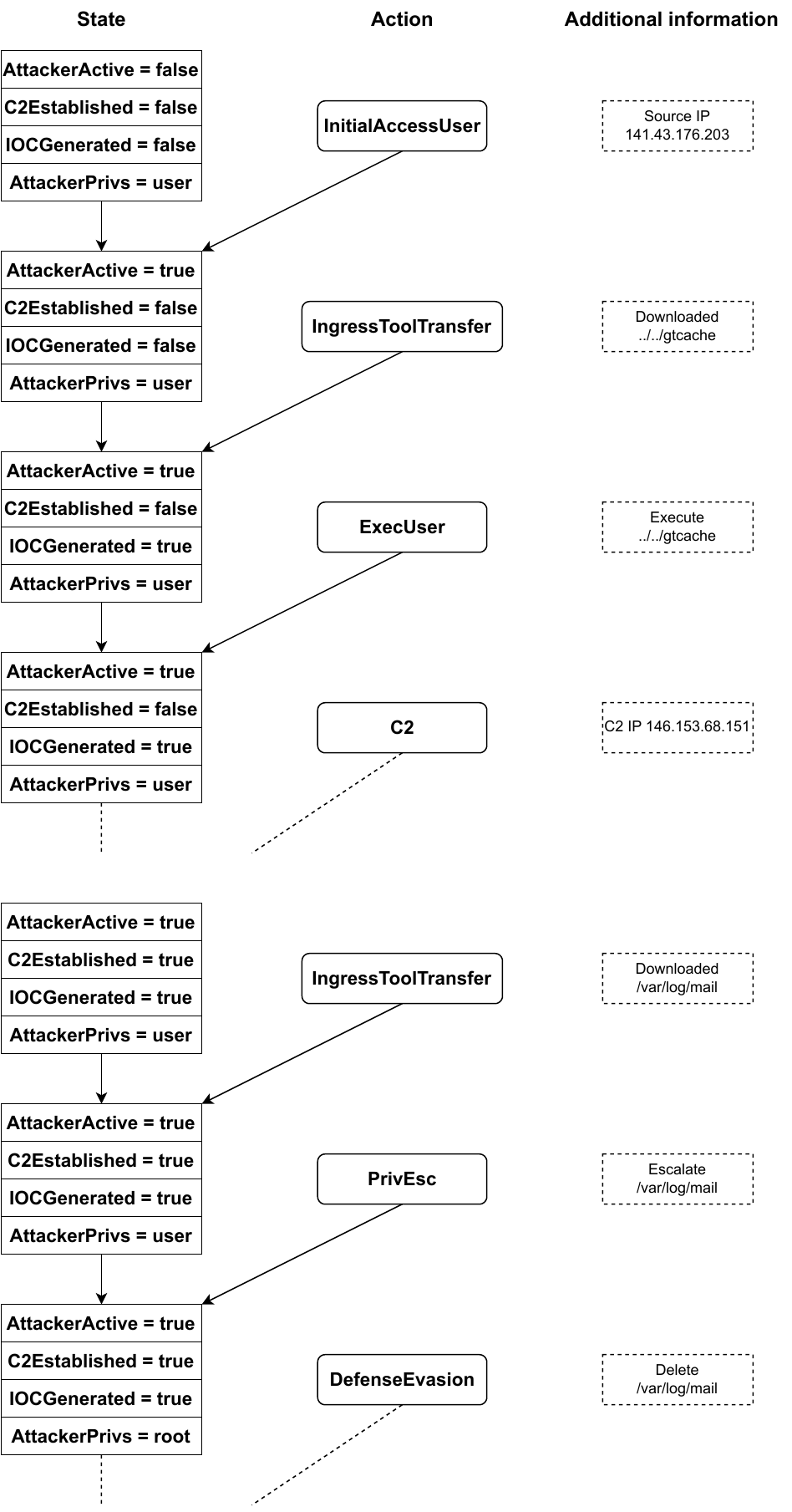}
    \caption{The state-action trajectory for the THEIA-2 attack}
    \label{fig:theia2}
\end{figure}

\begin{figure*}[ht]
\centering
    \includegraphics[width=0.7\linewidth]{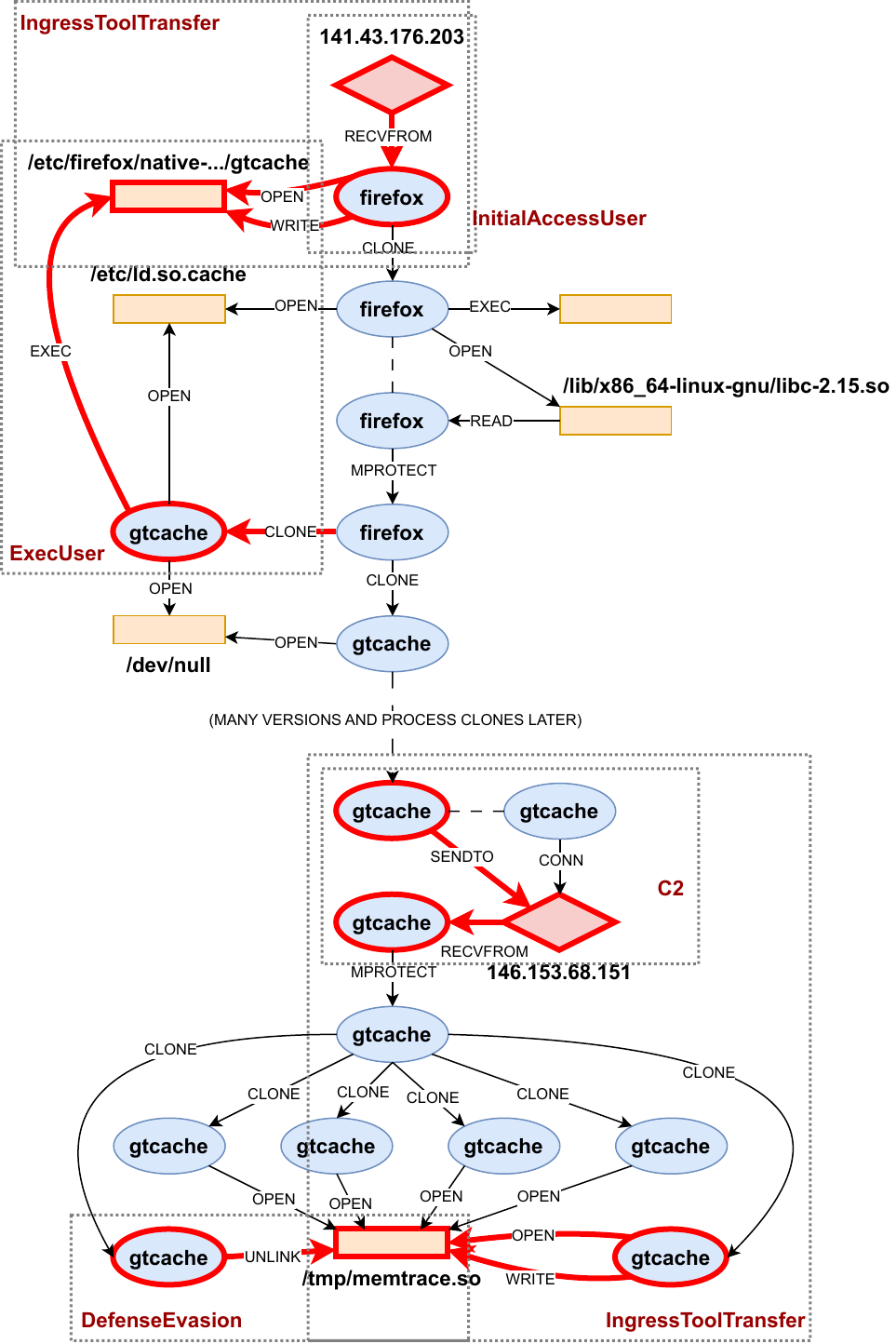}
    \caption{A small subgraph of the state-versioned provenance graph for the THEIA-2 attack shows the attacker's activity on the target system. The nodes and edges highlighted in red match the subgraph templates for attacker actions in the MDP model. The dotted boxes surrounding those subgraphs indicate the action that was identified using subgraph isomorphism}
    \label{fig:theia2_prov}
\end{figure*}

The THEIA-2 attacker started with user-level privileges by exploiting the Firefox browser. The attacker downloaded the APT stage /etc/firefox/native-messaging-hosts/gtcache. The attacker executed this APT stage with user-level privileges. Subsequently, the attacker downloaded additional APT stages /var/log/wdev, /tmp/memtrace.so, and /var/log/mail. However, the attacker only elevated /var/log/mail to root privileges and deleted all the other files. Throughout the attack, the attacker communicated with 146.153.68.151 for command and control.Figure~\ref{fig:theia2} shows the THEIA-2 trajectory. Figure~\ref{fig:theia2_prov} shows a small provenance subgraph of the THEIA-2 attack and attacker actions obtained using subgraph isomorphism with the template graphs.

\end{document}